\documentclass[twocolumn,floatfix,showpacs,prd,aps,tightenlines]{revtex4}
\usepackage{amsmath,amssymb,amsfonts}
\usepackage{graphicx}
\usepackage{psfrag}
\usepackage{dcolumn}
\usepackage{bm}

\begin{document}

\title{Spin-orbit interactions in black-hole binaries}

\author{M. Campanelli}  \affiliation{Department of Physics and Astronomy,
and Center for Gravitational Wave Astronomy,
The University of Texas at Brownsville, Brownsville, Texas 78520}

\author{C. O. Lousto} \affiliation{Department of Physics and Astronomy,
and Center for Gravitational Wave Astronomy,
The University of Texas at Brownsville, Brownsville, Texas 78520}

\author{Y. Zlochower} \affiliation{Department of Physics and Astronomy,
and Center for Gravitational Wave Astronomy,
The University of Texas at Brownsville, Brownsville, Texas 78520}

\date{\today}

\begin{abstract}

We perform numerical simulations of  black-hole binaries
to study the exchange of spin and orbital angular momentum during the
last, highly nonlinear, stages of the coalescence process. To
calculate the transfer of angular momentum from orbital to spin, we start
with two quasi-circular configurations, one with initially
non-spinning black holes, the other with corotating black holes. In
both cases the binaries complete almost two orbits before merging.
We find that, during these last orbits, the specific spin ($a/m$) of each horizon increases
by only $0.012$ for the initially non-spinning configuration, and
by only $0.006$ for the initially corotating configuration. By
contrast, the corotation value for the specific spin should
increase from $0.1$ at the initial proper separation of $10M$
to $0.33$ when the proper separation is $5M$.
Thus the spin-orbit coupling is far too weak to tidally lock the
binary to a corotating state during the late-inspiral phase.
We also study the converse transfer from spin into
orbital motion. In this case, we start the simulations with parallel,
highly-spinning non-boosted black holes. As the collision
proceeds, the system acquires a non-head-on orbital motion, due to
spin-orbit coupling, that leads to the radiation of angular
momentum. We are able to accurately measure the energy and angular
momentum losses and model their dependence on the initial spins.

\end{abstract}

 \maketitle

\section{Introduction}\label{Sec:Intro}

With the advent of new methods to evolve black-hole binaries
for several orbits~\cite{Pretorius:2005gq,Campanelli:2005dd,Baker:2005vv,Campanelli:2006gf,Baker:2006yw,Pretorius:2006tp,Scheel:2006gg}
we can now pose questions
of physical interest about the details of their interactions and
the generation of gravitational radiation, and find answers that allow
us to better understand the astrophysical properties of black-hole binaries.

The coalescence of unequal-mass black-hole binaries with randomly
oriented individual spins is expected to be one of the strongest
astrophysical sources of gravitational radiation. The first numerical
simulations of non-spinning unequal-mass black-hole binaries, along
with the calculation of the merger kicks, were reported in
Refs.~\cite{Campanelli:2004zw, Herrmann:2006ks, Baker:2006vn}.

In a preceding paper~\cite{Campanelli:2006uy} we performed the first
fully-nonlinear simulations of highly spinning black-hole binaries
and showed that the spin-orbit coupling leads to
significant changes in the orbital dynamics, as well as  
gravitation radiation emitted, in the final stages of a
black-hole binary. In this paper we study in more detail the
mechanisms that transfer orbital angular momentum to spin and {\it vice versa}.

It is of astrophysical interest to know if tidal effects could lock
the spin of black-hole binaries to a corotating state,
i.e.\ a state where the spins of the individual horizons are aligned with orbital
angular momentum, and the spin magnitudes
are such that the  horizon frequency $\Omega_H$ ($\Omega_H$ is the angular
speed of locally-non-rotating observers as they pass through the horizon,
as seen by stationary observers at infinity)
 is nearly equal to
 the orbital frequency $\Omega_O$. For a Kerr hole with mass $m$ and 
specific spin $a/m$ 
the horizon frequency takes the value
\begin{equation}
\label{eg:OmegaH}
\Omega_H=\frac{1}{2m}\left(\frac{a/m}{1+\sqrt{1-(a/m)^2}}\right).
\end{equation}
See Ref.~\cite{Caudill:2006hw} for a first post-Newtonian (1PN)
correction to the corotation condition $\Omega_O = \Omega_H$.
Note that in this paper we consider both the spin (or more accurately,
the spin angular momentum) $S$
of a black hole, which has units of $M^2$,
and the specific-spin $a/m$ ($m$ is the horizon mass),
which is dimensionless. The two are related by $S=a m$.  
We will also make use of the following relationships between the spin,
specific spin, horizon frequency, and
irreducible mass ($m_{\mbox{\tiny IR}}$):
\begin{eqnarray}
  \label{eq:SOmegaH}
  S &=& \frac{4 m_{\mbox{\tiny IR}}^3 \Omega_H}{\sqrt{1 - 4 (m_{\mbox{\tiny IR}} \Omega_H)^2}}, \\
  \label{eq:aOmegaH}
  a/m &=& 4 m_{\mbox{\tiny IR}} \Omega_H \sqrt{1 - 4 (m_{\mbox{\tiny IR}} \Omega_H)^2},
\end{eqnarray}
where $16 \pi m_{\tiny \mbox{IR}}^2$ is the horizon surface area.

  There are two
competing scales in this problem: one is the time for tidal effects to
act to spin up (or down) the black holes to the orbital frequency and the
other is the rate of change of the orbital frequency due to
gravitational radiation.

In the early stages of the orbital motion of black-hole binaries, tidal
effects are small and can be described by Newtonian theory; these
effects decay strongly with the binary separation $r$ (as $\mu/r^3$,
with $\mu$ the reduced mass of the system). Tidal  
effects will be stronger as the holes get closer,
but this
will also increase the orbital frequency (roughly like
$\Omega_O\sim\sqrt{\mu/r^3}$). In order to be correctly quantified, these effects
need to be evaluated dynamically with a fully non-linear numerical
simulation of the final orbital stage of black-hole binaries. Despite
this, several computations of quasi-circular sequences of initial data
for black-hole binaries have assumed corotating spins from larger
separations down to the estimated location of the innermost stable circular orbit (ISCO)
~\cite{Blanchet:2001id,Grandclement02,Caudill:2006hw}. 
Corotation has a notable effect on the location of the ISCO 
in all these calculations;
fortunately, recent numerical evolutions of equal-mass black-hole binaries 
~\cite{Campanelli:2006gf,Baker:2006yw,Pretorius:2006tp} do not find 
any particular physical effect associated with the location of the ISCO.

Bildsten and Cutler \cite{Bildsten92} (see also \cite{Kochanek:1992wk})
studied neutron star (NS) and black hole (BH) systems.  They conclude that, for
the NS-NS binary the viscosity is too low to tidally
synchronize the stars as they inspiral, and that there is simply not enough time during
the gravitational radiation driven inspiral of NS-BH binaries to tidally lock the spin
of the neutron star to the orbital period. Additionally, when the NS-BH binaries
are close enough for the tidal effects to be strong, the neutron stars are 
disrupted in a few orbits. 
Since these arguments
do not strictly apply to black-hole binaries, it is interesting to
explore these tidal interactions dynamically with fully non-linear numerical techniques.
This spin-up of NS-NS binaries has also been analyzed in
Ref.~\cite{Marronetti:2003gk}.

We will study in Sec.~\ref{Sec:spinup} the tidal spin-up of two equal-mass,
initially non-spinning, black holes starting from a quasi-circular
orbit (as determined by 3PN trajectories) with nearly two orbits left
before merger, as well as a similar configuration with initially
corotating black holes. These equal-mass configurations involving the very last stages of the
inspiral should nearly maximize the tidal effects.

While the transfer of orbital angular momentum to the spin of the members
of a binary system is of astrophysical relevance, we can also study the
conversion of spin into orbital angular momentum and verify the
importance of its dynamical effects.
To this end, we evolve binaries consisting of non-boosted, equal-mass black holes,
with the black holes aligned along the $y$-axis and the individual
spins aligned along the $+z$-axis, and measure the radiated
energy and angular momentum.
This simple set-up clearly displays the
transfer of spin to orbital motion. 
For non-spinning holes this set-up produces
purely head-on collisions, leading to the radiation of
energy but {\it not} angular momentum.  For spinning
holes, the spin-orbit coupling causes the holes to acquire
non-head-on orbital motion, leading to the radiation of angular
momentum. Thus, an accurate measure of the angular momentum radiated
would allow us to estimate the correct magnitude of this effect.

The leading post-Newtonian spin-orbit and spin-spin corrections
to the acceleration of two equal-mass black holes (each of mass $m$),
 with spins $\vec{S}_1$ and
$\vec{S}_2$ perpendicular to the motion, in a nearly head-on collision
are:
\begin{eqnarray}
\vec{a}_{SO}&=&-\frac{16}{r^3}(\vec{v}_r\times(\vec{S}_1+\vec{S}_2)),\\
\vec{a}_{SS}&=&-\frac{6}{mr^4}(\vec{S}_1\cdot\vec{S}_2)\,\hat{r},
\end{eqnarray}
where $\vec r$ is the binary separation, $\vec v_r$ is the relative
velocity, and $SO$ and $SS$ denote spin-orbit and spin-spin coupling terms 
respectively.
While the spin-spin acceleration adds an attractive term along the
radial direction (for parallel spins; repulsive for anti-parallel), the
spin-orbit acceleration is perpendicular to both the mostly-radial
(along $\hat{r}$) head-on velocity and the spin, and hence leads to a
non-vanishing orbital component of the binary motion.

In Sec.~\ref{Sec:spin-orbit} we will study this nonlinear coupling
using fully non-linear numerical evolutions and show that our method is able to extract
meaningful results for this numerically demanding relatively small effect.

\section{Techniques}\label{Sec:techniques}

We use the Brandt-Br\"ugmann puncture approach~\cite{Brandt97b} along with the 
{\sc TwoPunctures}~\cite{Ansorg:2004ds} and {\sc BAM\_Elliptic}~\cite{cactus_web}
thorns to compute initial data.
In this approach the 3-metric on the initial slice has the form
$\gamma_{a b} = (\psi_{BL} + u)^4 \delta_{a b}$,
where $\psi_{BL}$ is the Brill-Lindquist conformal factor,
$\delta_{ab}$ is the Euclidean metric,
and $u$ is (at least) $C^2$ on the punctures.
The Brill-Lindquist conformal factor
is given by
$
\psi_{BL} = 1 + \sum_{i=1}^n m_i / (2 r_i),
$
where
$n$ is the total number of `punctures', $m_i$ is the mass
parameter of puncture $i$ ($m_i$ is {\em not} the horizon
mass associated with puncture $i$), and $r_i$ is the coordinate
distance to puncture $i$. In all cases below, we evolve data containing
only two punctures with equal puncture mass parameters, and we denote
this puncture mass parameter by $m_p$.
 We evolve these black-hole-binary data sets using the {\sc
LazEv}~\cite{Zlochower:2005bj} implementation of the moving puncture
approach~\cite{Campanelli:2005dd, Baker:2005vv}.  In our version of
the moving puncture approach~\cite{Campanelli:2005dd} we replace the
BSSN~\cite{Nakamura87,Shibata95, Baumgarte99} conformal exponent
$\phi$, which has logarithmic singularities at the punctures,
 with the initially
$C^4$ field $\chi
= \exp(-4\phi)$.  This new variable, along with the other BSSN
variables, will remain finite provided that one uses a suitable choice
for the gauge. An alternative approach uses standard finite
differencing of $\phi$~\cite{Baker:2005vv}. Note that both approaches
have been used successfully by several other
groups~\cite{Sperhake:2006cy, Herrmann:2006ks, Hannam:2006vv}.

We obtain accurate, convergent waveforms and horizon parameters
by evolving this system in
conjunction with a modified 1+log lapse, a modified Gamma-driver shift
condition~\cite{Alcubierre02a,Campanelli:2005dd}, and an initial lapse
$\alpha\sim\psi_{BL}^{-4}$.  The lapse and shift are evolved with
$(\partial_t - \beta^i \partial_i) \alpha = - 2 \alpha K$,
$\partial_t \beta^a = B^a$, and $\partial_t B^a = 3/4 \partial_t \tilde \Gamma^a - \eta B^a$.
These gauge conditions require careful treatment of $\chi$
near the puncture in order for the system to remain
stable~\cite{Campanelli:2005dd,Campanelli:2006gf}. In
Ref.~\cite{Gundlach:2006tw} it was
shown that this choice of gauge leads to a strongly hyperbolic
evolution system provided that the shift does not become too large.
For our version of the moving puncture approach, we find that the
product $\alpha \tilde A^{ij} \partial_j \phi$ initially has to be $C^4$ on the
puncture. In the spinning case, $\tilde A^{ij}$ is $O(r^3)$ on the
puncture, thus requiring that $\alpha \propto r^3$ to
maintain differentiability. We therefore choose an initial lapse
$\alpha(t=0) = 2/(1+\psi_{BL}^{4})$ which is $O(r^4)$ and $C^4$ on the
puncture and reproduces the isotropic Schwarzschild lapse at large
distances from the horizons.  The initial values of $\beta^i$ and $B^i$
are set to zero.

Hannam~et.~al.~\cite{Hannam:2006vv} examine the smoothness of the evolved fields at late
times at the puncture. They find that, in the case of Schwarzschild,
 $\chi$ transitions from an initially $C^4$ 
field to a $C^2$ field at late times. Although we require that the fields are
initially $C^4$, this late-time drop in smoothness does not
appear to leak out of the horizon (which is consistent with the
analysis in~\cite{Hannam:2006vv}).

We use a `multiple transition' fisheye transformation~\cite{Campanelli:2006gf}
to push the boundaries to $100M$, while maintaining a resolution
of up to $M/40$ in the central region,  for the head-on runs; for orbital
runs we push
 the boundaries to $215M$, with a central resolution of up to $M/31.5$.

We use a locally modified version of the {\sc AHFinderDirect}
 thorn~\cite{Thornburg2003:AH-finding}
to find the individual apparent horizon, and
we measure the angular momentum of the horizons using
our implementation of the isolated
horizon algorithm detailed in~\cite{Dreyer02a}.
(The isolated horizon formalism used here is extended to the
dynamical case in~\cite{Ashtekar03a}.) In the isolated
horizon approach, the horizon spin is given by
\begin{equation}\label{isolatedspin}
    S = \frac{1}{8\pi}\oint_{AH} (\varphi^aR^bK_{ab})\,\textrm{d}^2V .
\end{equation}
where $\varphi^a$ is an approximate Killing vector on the horizon,
$K_{ab}$ is the extrinsic curvature,  $d^2V$ is the
natural area element intrinsic to the horizon, and $R^a$ is the
outward pointing unit vector normal to the horizon on the 3D slice.
(See appendix~\ref{app:isolated_v_cir} for a comparison of the
isolated horizon technique for finding the horizon spin 
to the more common horizon circumference technique.)
 We confirm that our
version of the algorithm produces the correct spin for various
coordinate distortions of Kerr, as well as for Bowen-York binary
initial data (both with and without linear momentum).

We measure the
proper distance between the horizons by shooting geodesics from the
origin to a horizon, and then finding the geodesic with the minimum
proper distance as described in Refs.~\cite{Koppitz04,Diener:2005mg} (the proper distance
between the two holes is then double this distance). This technique will
work in cases where there is (at least) Pi-symmetry on the
equatorial plane and reflection symmetry across the plane.

\section{Tidal spin-up of black-hole binaries}\label{Sec:spinup}

The spin of the individual horizons in a black-hole binary will become
tidally locked to the orbital frequency if the non-linear spin-orbit
coupling is sufficiently strong
that the timescale for tidally induced spin-up (or down) is smaller than
the timescale for change in the orbital period. In order to obtain an accurate
measure of the strength of this coupling we examine the cases of
initially non-spinning and initially corotating  close binaries, denoted
by S0 and SC respectively,
where the binary separation is small enough that the spin-orbit coupling is
large, but large enough that the binaries complete at least $\sim1.75$
orbits before merging. The initial data parameters for these
configuration are provided in Table~\ref{table:spinupID} and have
been obtained by finding quasi-circular orbits with the 3PN equations
of motion.
We originally studied the S0 non-spinning configuration in
Ref.~\cite{Campanelli:2006uy}. While in~\cite{Campanelli:2006uy}
we were interested in the spin of the remnant horizon and plunge times
as a function of the initial spin of the individual horizons, here we
are interested only in the pre-merger spin-up of the individual
horizons. It is then crucial to have an appropriate measure
of the individual spin of the holes. 
We employ the isolated horizon
spin algorithm detailed in~\cite{Dreyer02a} (see Eq.~(\ref{isolatedspin})). 
Strictly speaking, isolated horizon techniques are only valid for
non-interacting horizons. However, they do provide an excellent
approximation to the horizon spin even in the case of relatively
close binaries. We confirmed that our implementation gives accurate
evaluations of the horizon spins for closely separated Bowen-York
binary data. For this test we constructed a sequence of initial data using
the {\sc TwoPunctures} thorn~\cite{Ansorg:2004ds} with fixed orbital angular momentum
$L=0.876 M^2$ (the value for the S0 case above) and ADM mass of $(1.0000\pm0.0005)M$. We set the spin of the
individual holes (aligned along the $+z$-axis) to $S=0M^2$, $S=0.001M^2$, and $S=0.0001M^2$, and measured
the spins of the horizons with our implementation of the isolated horizon
algorithm. The results of these calculations are summarized in 
Table~\ref{table:BYspin}. From the table we see that spins as low as
$10^{-4}M^2$ can be measured with an accuracy of better than $20\%$ 
even when the holes are separated by a proper distance of
$l\sim 3M$, and can be measured with an accuracy of $1\%$ or better when the holes are separated by
$l\geq 4.6M$. Interestingly, accurate measurement
of this spin can be obtain even with coordinate separations as small as $0.5M$.

\begin{table}
\caption{Initial data for quasi-circular, equal-mass black-hole binaries.
The binaries have an ADM mass of $(1.0000\pm0.0005)M$, with 
orbital frequency $M\Omega$ fixed to $0.0500$, total angular momentum (orbital + spin)
given by $(0,0,J)$, and proper horizon separation $l$.
 The punctures
are located at $(0,\pm Y,0)$, with mass parameter $m_p$, momentum
$(\mp P,0,0)$, spin angular momentum $(0,0,S)$, and specific spin
$a/m$ ($m$ is the horizon mass). 
}
\begin{ruledtabular}
\begin{tabular}{llllllll}\label{table:spinupID}
Name & $S/M^2$ & $Y/M$ & $P/M$ & $J/M^2$ & $a/m$ & $m_p/M$ & $l/M$\\
\hline
S0  & 0.000  & 3.280  & 0.1336  & 0.876  & 0.0000    &  0.4848   & 10.01 \\
SC & 0.025757&3.2534 & 0.1330& 0.917  & 0.1001    &  0.4831   & 9.93 \\
\end{tabular}
\end{ruledtabular}
\end{table}

\begin{table}
\caption{The isolated horizon spin of a sequence of Bowen-York data with
given spins and ADM mass $(1.0000\pm0.0005)M$. Here $(\pm x,0,0)$ are the coordinate locations of the
punctures, $S$ is the measured spin (the Bowen-York spin parameter is given
in parentheses), and $l$ is the proper distance between horizons. In all cases
the linear momentum is given by $\vec P = (0,\pm 0.8764M^2/(2 x),0)$
 (to reproduce the S0 orbital angular momentum),
 and the spins are aligned with the $z$-axis. }
\begin{ruledtabular}
\begin{tabular}{lllll}\label{table:BYspin}
$x/M$ & $S/M^2(10^{-3})$ & $S/M^2(10^{-4})$ & $S/M^2(0)$ & $l/M$\\
\hline
$3.280$ & $1.000\times10^{-3}$ & $1.003\times10^{-4}$ & $3.0\times10^{-7}$ &$10.0$\\ 
$3.0$ & $1.011\times10^{-3}$ & $1.011\times10^{-4}$ & $1.1\times10^{-6}$ &$9.4$\\ 
$2.0$ & $9.998\times10^{-4}$ & $9.97\times10^{-5}$ & $3.4\times10^{-7}$ &$7.0$\\ 
$1.0$ & $1.001\times10^{-3}$ & $1.011\times10^{-4}$ & $1.13\times10^{-6}$ &$4.6$\\ 
$0.5$ & $1.01\times10^{-3}$ & $1.11\times10^{-4}$ & $1.08\times10^{-5}$ &$3.4$\\ 
$0.25$ & $1.02\times10^{-3}$ & $1.17\times10^{-4}$ & $1.7\times10^{-5}$ &$3.0$\\ 
\end{tabular}
\end{ruledtabular}
\end{table}

\begin{figure}
\begin{center}
\includegraphics[width=3.3in]{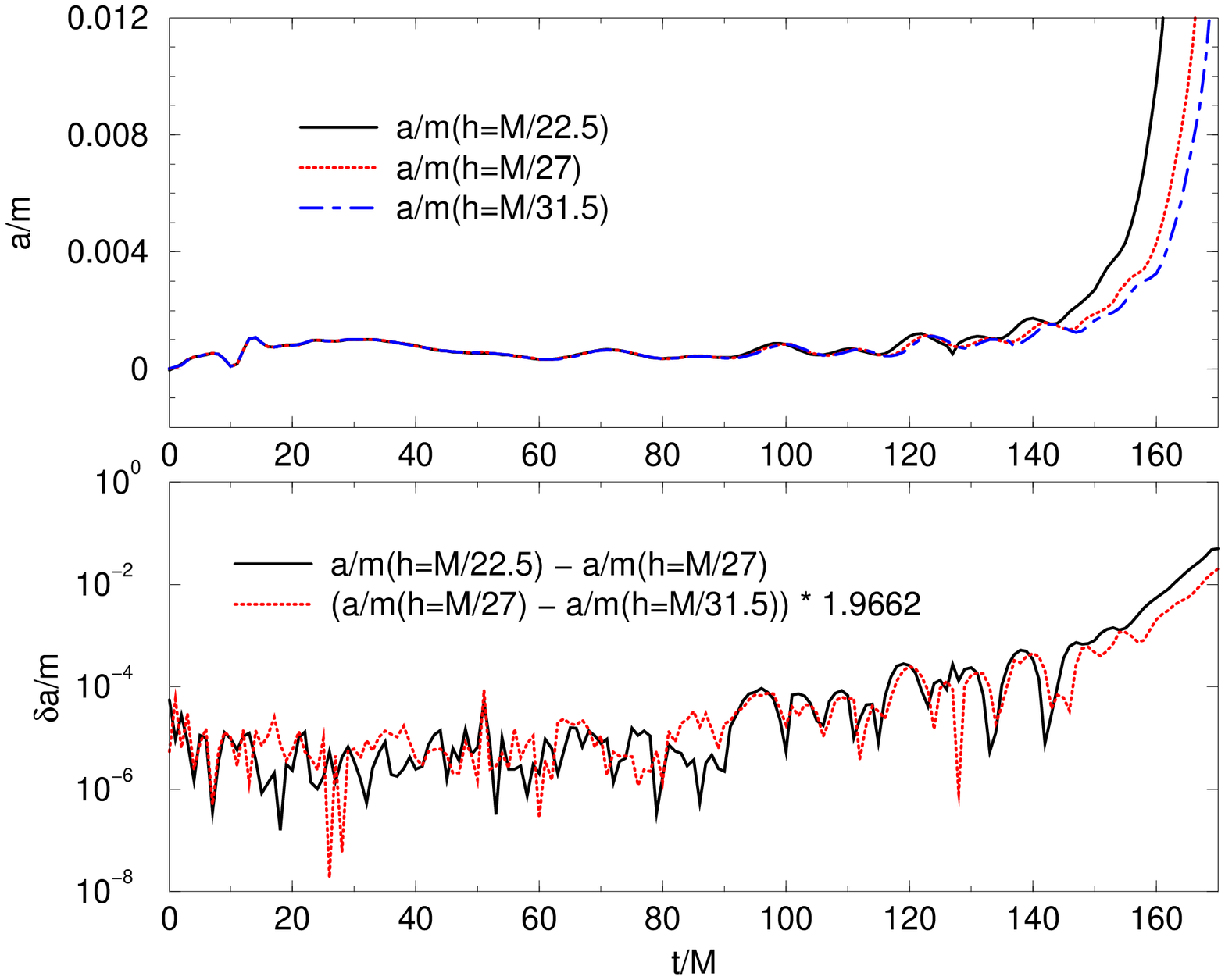}
\caption{The measured instantaneous specific spin of the individual horizons, for the
S0 configuration,  starting from vanishing spin at $t=0$
and up to the merger time. The top panel shows $a/m$ versus time for three resolutions
with grid-spacings $h=M/22.5$, $h=M/27$, and $h=M/31.5$,
while
the lower panel shows the differences in $a/m$ between the low and medium
resolutions and the medium and high resolutions (the latter rescaled by
1.9662 to demonstrate third-order convergence).
The differences between the low and medium resolutions and medium and high resolutions become large at
late times due to the lower resolution runs merging sooner. 
The curves in the top panel have been cut off at the approximate merger value of $a/m$
(which is independent of resolution).}
\label{fig:bs00_spinup}
\end{center}
\end{figure}

\begin{figure}
\begin{center}
\includegraphics[width=3.3in]{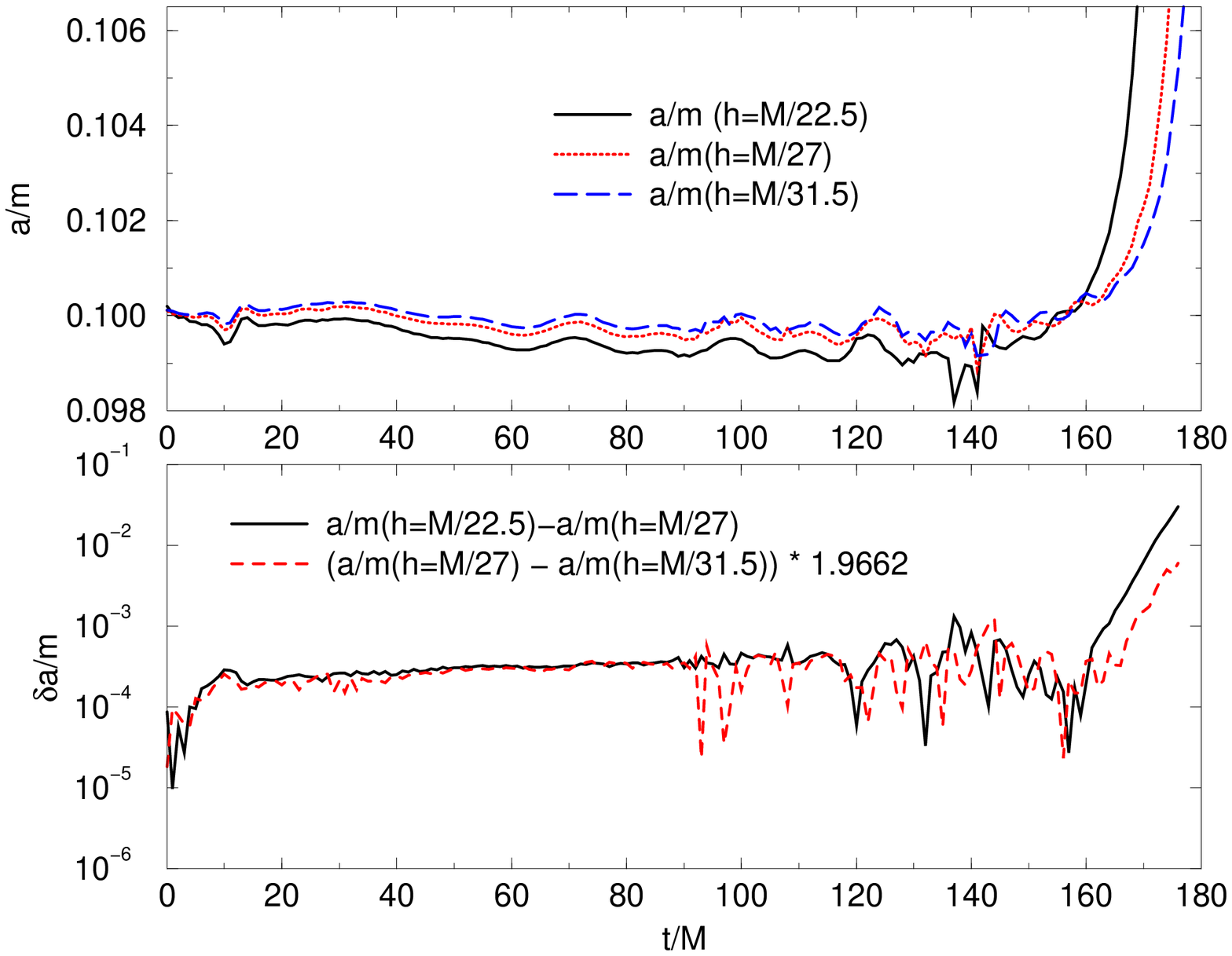}
\caption{The measured instantaneous specific spin of the individual horizons, for the
SC configuration,  starting from $a/m=0.1$ at $t=0$
and up to the merger time. The top panel shows $a/m$ versus time for three resolutions
with grid-spacings $h=M/22.5$, $h=M/27$, and $h=M/31.5$,
while
the lower panel shows the differences in $a/m$ between the low and medium
resolutions and the medium and high resolutions (the latter rescaled by
1.9662 to demonstrate third-order convergence).
The differences between the low and medium resolutions and the medium and high resolutions become large at
late times due to the lower resolution runs merging sooner.
The curves in the top panel have been cut off at the approximate merger value of $a/m$ 
(which is independent of resolution).}
\label{fig:bs0.1_spinup}
\end{center}
\end{figure}

\begin{figure}
\begin{center}
\includegraphics[width=3.3in]{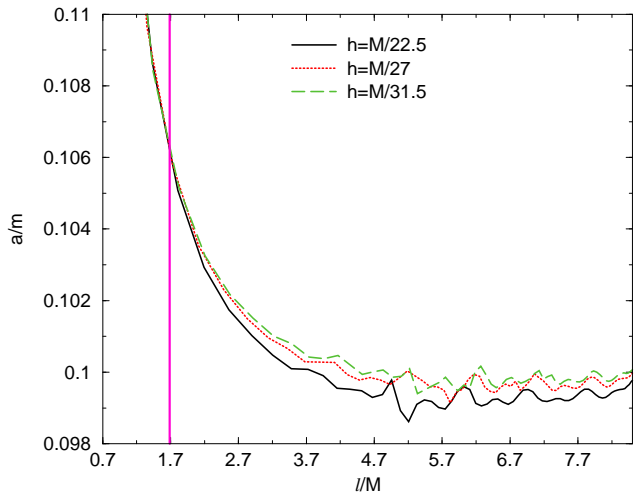}
\caption{The specific spin $a/m$ for the SC configuration versus proper binary
separation $l$ for three resolutions with grid-spacings $h=M/22.5$, $h=M/27$, and $h=M/31.5$.
Note that the large late-time phase errors seen in $a/m$ versus time are
not present in this plot. The vertical line shows the proper distance at merger.
}
\label{fig:bs0.1_avl}
\end{center}
\end{figure}

We evolved the S0 and SC configurations with 3 different resolutions, 
$h=M/22.5$, $h=M/27$, and $h=M/31.5$, where $h$ is the central
grid-spacing, and calculated the horizon spins after every $1M$ of evolution.
The results of measuring the spin of the individual holes during the
orbital motion until the formation of a common apparent horizon (CAH) (See Table
\ref{table:spinup-results}) for the S0 configuration are displayed in
Fig.~\ref{fig:bs00_spinup}, while the results for the SC configuration are
displayed in Fig.~\ref{fig:bs0.1_spinup}.
The upper panel in both figures shows the specific spin of the individual
holes for the three resolutions. We observe that in both cases the specific-spin
increase during the evolution is small ($0.012$ for S0 and
$0.006$ for SC).
The lower panels show essentially third-order convergence as expected
from our use of third-order-accurate interpolators with {\sc AHFinderDirect}.
The late-time error in the S0 specific spin for the lowest resolution run
appears to show a very large error compared to the actual value of the specific spin.
This `error' is actually a phase error resulting from the lower resolution
runs merging sooner.
However, in a specific spin versus proper separation plot, the curves
from all three resolutions lie on top of each other 
(see Fig.~\ref{fig:bs0.1_avl}) at late-times (i.e.\ at small proper distance).
This dichotomy, between the late-time convergence break-down of $a/m$
versus $t$ and the excellent late-time agreement between resolutions
in $a/m$ versus $l$, is similar to the results previously seen in the
waveforms of inspiraling
binaries~\cite{Baker:2002qf,Baker:2006yw,Campanelli:2006uy}.
In that case it was observed that $\psi_4(t)$
lost convergence due to large phase errors induced by the lower resolution
runs merging sooner, while the phase-corrected
$\psi_4$ showed excellent agreement between resolutions.
In this case the `phase error' is in $l(t)$, and by plotting $a/m$ versus
$l$ we effectively removed the phase error.

\begin{table}
\caption{Merger time, $T_{CAH}/M$, for the S0 and SC configurations versus resolution.
Horizon searches were performed every $0.3M$ and $0.2M$ for the S0 and 
SC configuration respectively. }
\begin{ruledtabular}
\begin{tabular}{lll}\label{table:spinup-results}
${\rm resolution}$ & S0 & SC\\
\hline
M/22.5 & $160.7\pm0.3$ & $168.6\pm0.2$\\
M/27 & $166.0\pm0.3$ & $174.2\pm0.2$\\
M/31.5 & $168.3\pm0.3$ & $176.6\pm0.2$\\
extrapolation & $172\pm2$ & $179\pm2$\\
\end{tabular}
\end{ruledtabular}
\end{table}

\begin{figure}
\begin{center}
\includegraphics[width=3.3in]{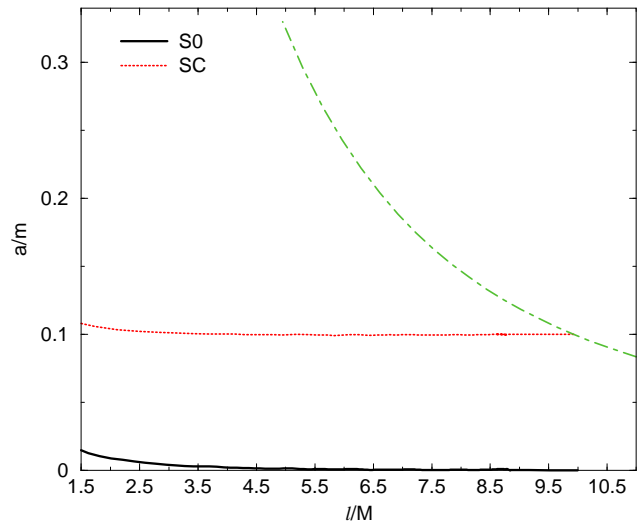}
\caption{The spin-up of the horizons versus proper binary separation.
The continuous line shows $a/m$ for the S0 configuration, the
dotted line shows $a/m$ for the SC configuration (initially
corotating), and the dot-dash line shows the corotation value
for the spin. Note that the S0 configuration shows a slightly
larger spin-up and that both spin-ups are much smaller than what is required
to tidally lock the binary.
}
\label{fig:a_v_l}
\end{center}
\end{figure}

The values that we obtained for the spin-up of the binary holes
are much smaller than those expected for a corotation
state
\footnote{Very rarely our isolated horizon routine fails to find the correct
approximate Killing vector and gives a very inaccurate value for the spin.
When this happens (we observes this no more than once per run for these orbital runs
and never for the head-on runs below) we drop the data point.
}.  In order to visualize this we have displayed in
Fig.~\ref{fig:a_v_l} the measured spin during the orbital motion, this
time against the proper separation of the apparent horizons.
The S0 configuration forms a common apparent horizon (CAH) at a proper separation
of $l_{CAH} = 1.75M$, while the SC configuration forms
a common apparent horizon at a proper separation of
$l_{CAH} = 1.69M$.
In order to have a measure of what the corresponding corotation
spin is, we considered the family of corotating binaries in quasi-equilibrium
 of Ref.~\cite{Grandclement02} (we have used Eq.~(\ref{eq:aOmegaH}) to convert the
$\Omega_H$ and $m_i$ provided in~\cite{Grandclement02} into
$a/m$). From the plot we see that the SC is
initially corotating, but the specific spin remains flat for most
of the run. Thus the spin-orbit coupling is far too small to keep
this system tidally locked. It is interesting to note that the 
spin-up for the S0 configuration is slightly larger than the spin-up for the SC
configuration. Both configuration show essentially no spin-up above
proper separations of $\sim 4M$.

The relatively low `spin up' that we measure during the last couple of orbits
prior to the merger indicates that the time-scale for tidal locking
is much larger than the time-scale of orbital decay due to gravitational 
radiation emission. We also note that  in the leading relevant
post-Newtonian order, the change of the spin, averaged over a circular orbit,
which is given
by~\cite{Kidder:1995zr}
\begin{eqnarray}\label{Sdot}
\dot{\vec{S}}_i&=&\frac{1}{2r^3}\left\{(\vec{L}_N\times\vec{S}_i)
(4+3\frac{m_j}{m_i})+\vec{S}_j\times\vec{S}_i\right.\nonumber\\
&&\left.-3(\hat{L}_N\cdot\vec{S}_j)
\hat{L}_N\times\vec{S}_i\right\} \quad(j\neq i),
\end{eqnarray}
vanishes for all configurations with the spins aligned with the
orbital angular momentum. In Eq.~(\ref{Sdot})
$\vec{L}_N$ is the (Newtonian) orbital angular momentum, $\vec{S}_i$
is the spin of black hole $i$, $m_i$ is its mass, and $j$ denotes the other black hole.
Will~\cite{Will:2005sn} also points out
that the radiation reaction interaction does not modify the spin up to
3.5PN order. First effects of the flattening of the bodies should
actually be seen at 4.5PN order.

\section{Spin-orbit couplings in black-hole binaries}\label{Sec:spin-orbit}

We continue our study of the relevance of spin-orbit coupling to the
orbital dynamics of black-hole binaries by choosing initially head-on
(vanishing transverse linear momentum)
configurations that single out the converse transfer of spin into
orbital angular momentum.  Purely head-on collisions do not radiate
angular momentum to infinity. Thus, in these configurations, the
radiated angular momentum is an indicator of the extent to which  these
systems deviate from head-on motion (i.e.\ the extent to which they
acquire transverse motion). The energy and angular momentum radiated are
functions of the spins of the two horizons, the initial inward momentum of
each hole, and the proper separation.
To simplify the analysis, we choose head-on binaries with zero initial
linear momentum and with a fixed initial proper separation of $12.24M$. All
spins are aligned along the $z$-axis, with the binaries initially
separated
along the $y$-axis. We report the initial data parameters in
Table~\ref{table:HeadonID}.

\begin{table}
\caption{Initial data for non-boosted black-hole binaries. The binaries
have an ADM mass of $(1.0000\pm0.00005)M$, with total angular momentum $(0,0,J)$,
and proper horizon separation $l$. The punctures are located at
$(0,\pm Y,0)$, with mass parameter $m_p$, spins angular momentum $(0,0,S)$,
and specific spin $a/m$ (here $m$ is the horizon mass).}
\begin{ruledtabular}
\begin{tabular}{llllllll}\label{table:HeadonID}
Name & $S/M^2$ & $Y/M$ & $P/M$ & $J/M^2$ & $a/m$ & $m_p/M$ & $l/M$\\
\hline
HS++  & 0.200  & 4.000  & 0.000 & 0.4  &  0.7581    &   0.3444  & 12.24 \\
MS++  & 0.100  & 4.2422  & 0.000 & 0.2  & 0.3775    &  0.469755  & 12.24 \\
0S++  & 0.000  & 4.2924  & 0.000 & 0.0  & 0.0000    &  0.5000   & 12.24 
\end{tabular}
\end{ruledtabular}
\end{table}

\begin{figure}
\begin{center}
\includegraphics[width=3.3in]{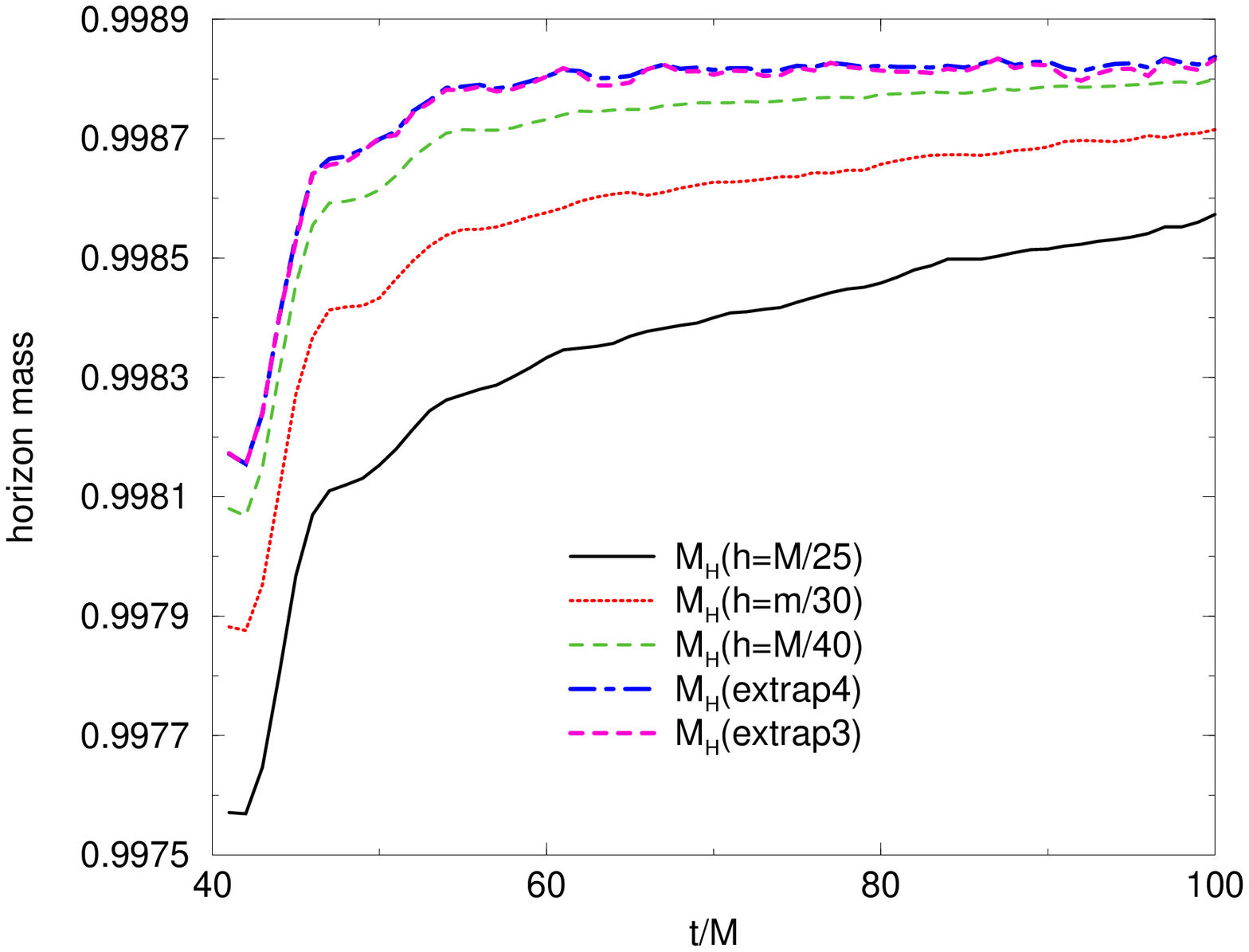}\\
\includegraphics[width=3.3in]{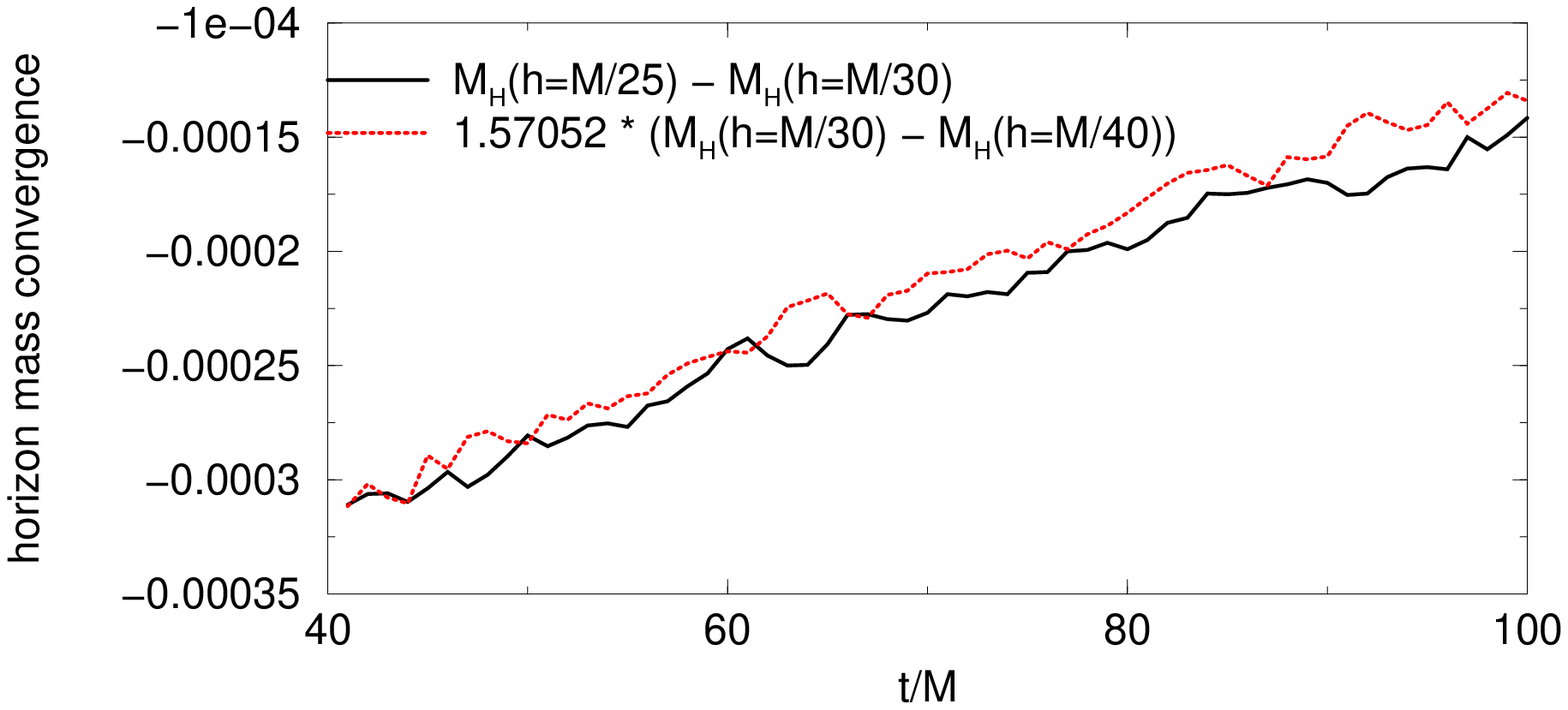}
\caption{The final remnant horizon mass versus time for the HS++ configuration.
The top panel shows the horizon mass for the three resolutions
with grid-spacings $h=M/25$, $h=M/30$, and $h=M/40$, and 
the two Richardson extrapolations based on  leading third-order
and leading fourth-order errors. The bottom panel shows the differences
in the horizon mass for the $h=M/25$ and $h=M/30$ runs as well as the differences
in the mass for the $h=M/30$ and $h=M/40$ runs. This latter difference is
rescaled by 1.57052 to demonstrate 3.7-order convergence.
The accurate extrapolations (as evident by the agreement of the two
extrapolations) to infinite resolution allows for a
sufficiently precise determination of the radiated energy to model
its dependence on the initial spins.
}
\label{fig:VHS_PP_M_ex}
\end{center}
\end{figure}

\begin{figure}
\begin{center}
\includegraphics[width=3.3in]{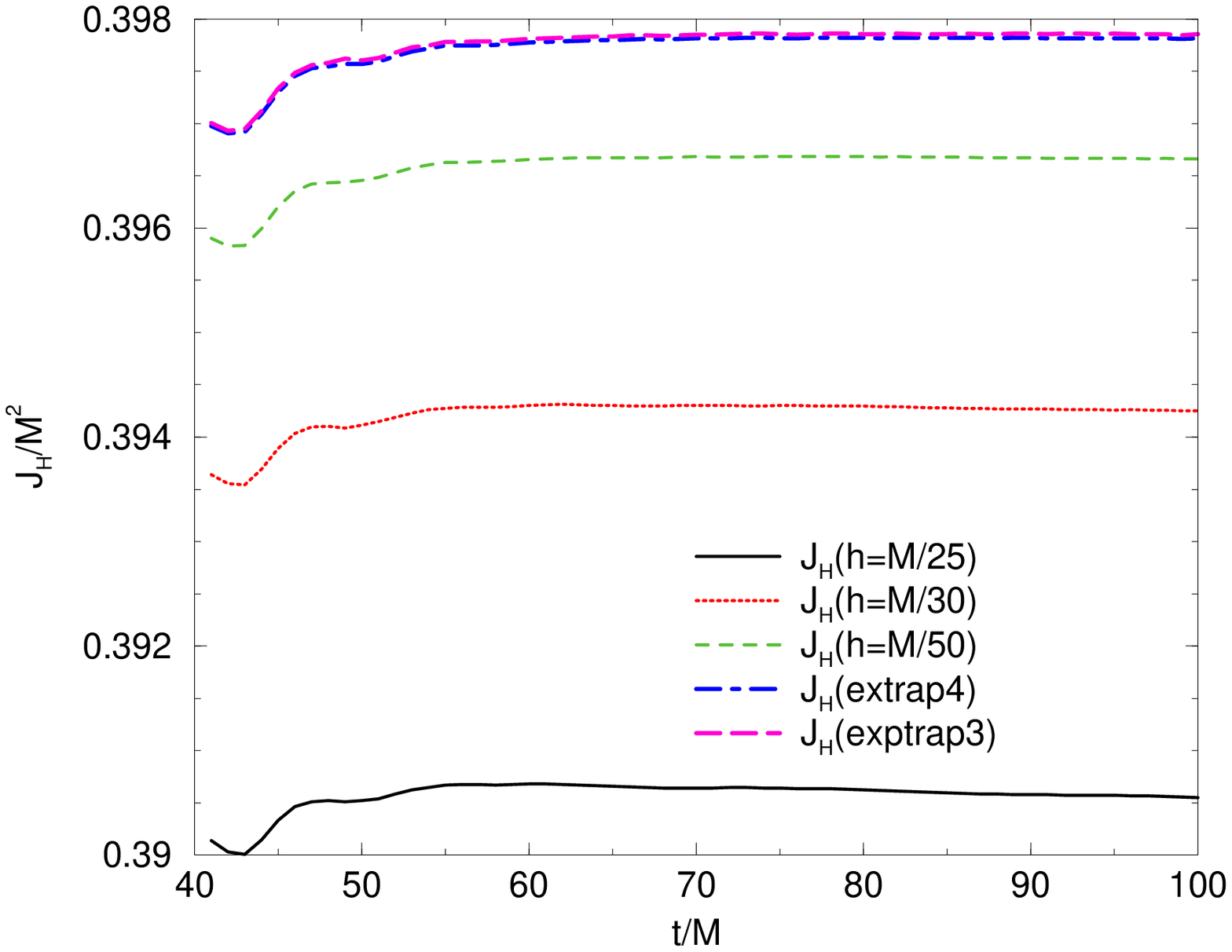}
\includegraphics[width=3.3in]{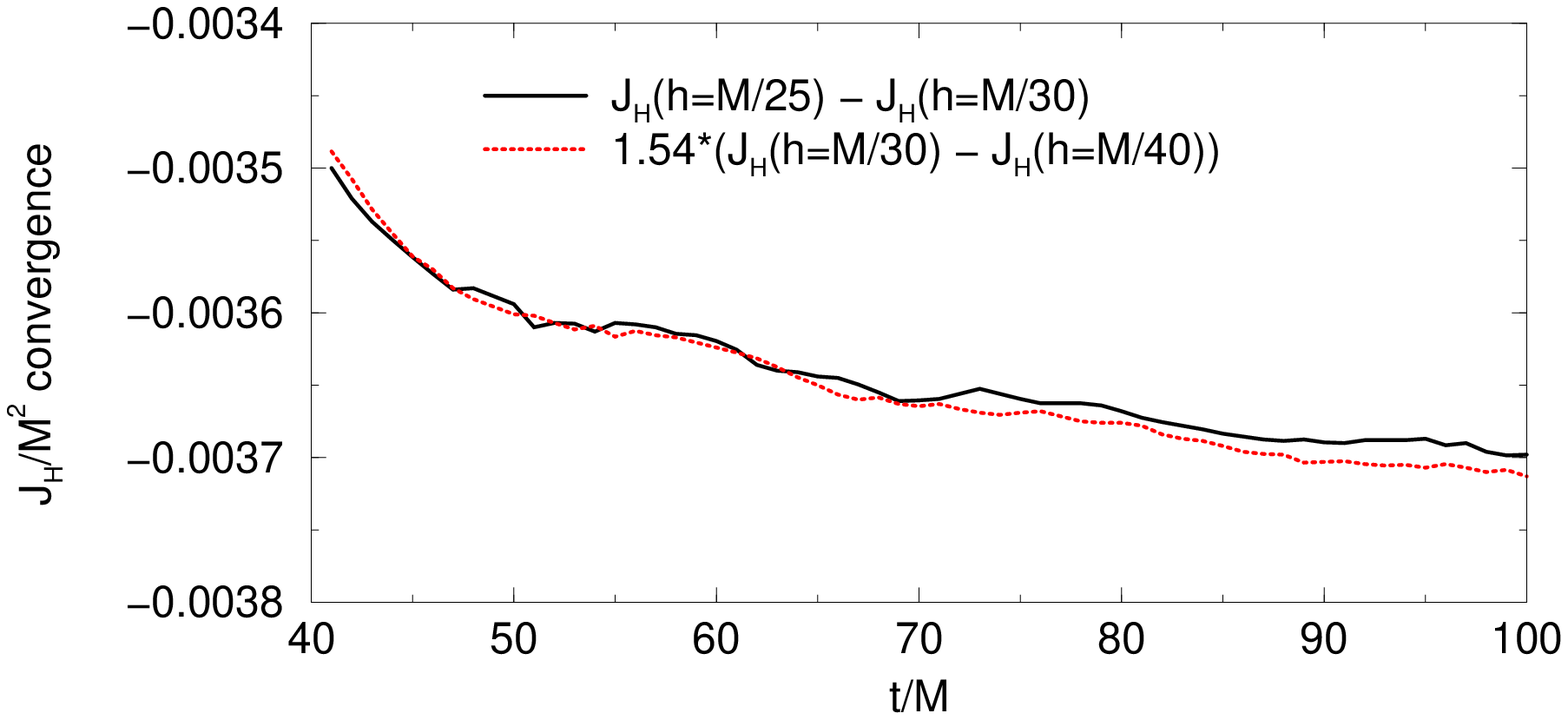}
\caption{The angular momentum of the final remnant black hole for the
HS++ configuration measured by
the isolated horizon method.
The top panel shows the horizon spin for the three resolutions
with grid-spacing $h=M/25$, $h=M/30$, and $h=M/40$, as well as
the two Richardson extrapolations based on  leading third-order
and leading fourth-order errors. The bottom panel shows the differences
in the horizon spin for the $h=M/25$ and $h=M/30$ runs as well as the differences
in the spin for the $h=M/30$ and $h=M/40$ runs. This latter difference is
rescaled by 1.54 to demonstrate 4th-order convergence.
The  accurate extrapolations (as evident by the agreement of the two
extrapolations) to infinite resolution allows for a sufficiently precise
determination of the radiated angular momentum to model its dependence
on the initial spins.
}
\label{fig:VHS_PP_J_ex}
\end{center}
\end{figure}

We measure the radiated energy and angular momentum by solving for
the final hole remnant horizon mass and spin, and then subtracting
these quantities from the ADM mass and angular momentum (as determined from the
initial data). To avoid confusion with the initial black-hole mass
and spins, we denote the mass and spin of the final remnant
horizon with $M_{H}$ and $J_{H}$ respectively. The radiated
mass and angular momentum then are:
\begin{eqnarray}
  E_{rad} &=& M_{ADM} - M_{H},\\
  J_{rad} &=& J_{ADM} - J_{H},
\end{eqnarray}
where in these head-on cases $J_{ADM}$ is the sum of the spins
of the two initial black holes. For notational convenience
we introduce the symbol $j_0$ defined by
\begin{equation}
j_0 = \frac{J_{ADM}}{M_{ADM}^2}.
\end{equation}
In previous papers~\cite{Campanelli:2006gf, Campanelli:2006uy}
we found that the horizon provided a very accurate
measurement of these quantities. Additionally, when using isolated
horizon techniques, we can obtain the horizon parameters accurately
within $30M$ after the merger.
In all cases the radiated energy and angular momentum are very small.
In order to accurately obtain these numbers we need to evolve all
configurations at three different resolutions (see Table~\ref{table:Headon-results}
for the grid-spacings) and then use a
Richardson extrapolation of the radiated energy and angular momentum.
The measured convergence rates for the radiated angular momentum are
$3.9$ and $3.7$ for the HS++ and MS++ configurations respectively, while the
measured convergence rates for the radiated energy are $4.0$, 
$3.7$, and $3.9$ for the HS++, MS++, and 0S++ configurations respectively.
The lower convergence rates for the MS++ configuration
may simply indicate only that higher resolution is needed. Alternatively it is possible that
small third-order errors (possibly from the interpolators) are present in all
configurations, but are only observed in the MS++ configuration due
to its finer effective resolution. (The effective grid-spacing 
is $h_{eff} = h/m_p$, where $h$ is the actual grid-spacing
and $m_p$ is the puncture mass parameter (see~\cite{Campanelli:2006uy} for details)).
 For the Richardson
extrapolation we use both the forms $f(h) = f_{\infty} + e\,h^4+ g\,h^5$
and $f(h) = f_{\infty} + d\,h^3 + e\,h^4$, and, using results from the
three resolutions, solve for $f_{\infty}$.
 We take the difference
between these two extrapolation to be indicative of the error in our extrapolated numbers.
In Figures~\ref{fig:VHS_PP_M_ex}~and~\ref{fig:VHS_PP_J_ex} we show the
HS++ final remnant black-hole horizon mass and spin
versus time for three resolutions, as well as the two extrapolations of
these quantities and convergence plots. It is evident from the very close
agreement between the two extrapolations that we obtain highly accurate
measurements of both the horizon mass and spin, and consequently,
the radiated mass and angular momentum.

In Table~\ref{table:Headon-results} we summarize the results for these
configurations. Note that the errors reported in the table for
a given resolution are of the
uncertainty in measurement of  $J_{H}$ and $M_{H}$ at that resolution (i.e.\ a measure of
the flatness of the late-time plateaus of $J_{H}$ and $M_{H}$). Thus the error
interval reported for e.g.\ $J$ radiated at a resolution of $M/25$ will not 
necessarily contain the final extrapolated value. On the other hand,
the error intervals for the extrapolated values are our best estimate for the true value of
each quantity. A fit of the Richardson-extrapolated values of
the radiated angular momentum to the initial spins of the individual black holes
has the form:
\begin{eqnarray}
\label{eq:JradVJ}
\frac{J_{rad}}{M^2}&=& (4.8\pm0.1)\times10^{-3}j_0+\nonumber\\ 
&&(4.0\pm0.9)\times 10^{-3} j_0^3 + 
 O\left(j_0^5\right),
\end{eqnarray}
where $j_0 = 2 S/M_{ADM}^2$ (see Table~\ref{table:HeadonID}).
The linear term is the dominant one  for these configurations (the HS++ configuration
radiates $\sim20\%$ more momentum than expected from the linear behavior alone).
Equation~(\ref{eq:JradVJ}) contains only odd powers of $j_0$ since the magnitude of
$J_{rad}$ must be independent of $j_0$ (i.e.\ the physics
would not change if the two spins were aligned along the $-z$-axis),
 and the sign of $J_{rad}$ must match the sign of  $j_0$. 

\begin{widetext}

\begin{table}
\caption{The radiated mass and angular momentum as well as the
merger time ($T_{CAH}$) and final specific spin for the head-on configurations.
 The reported
error intervals in $T_{CAH}$ are a function of how often a horizon
search was performed. In all cases $T_{CAH}$ does not appear to
vary with resolution.}
\begin{ruledtabular}
\begin{tabular}{llllll}\label{table:Headon-results}
{\rm Case} & ${\rm resolution}$ & $E_{rad}/M_{ADM}$ & $J_{rad}/J_{ADM}$ & $T_{CAH}/M$ & $a/M_{H}$\\
\hline
HS++  	& M/25 & $(0.16\pm0.02)\%$ & $(2.37\pm0.02)\%$ & $36.7\pm0.2$ & $0.3918\pm0.002$\\
      	&   M/30 & $(0.136\pm0.009)\%$ & $(1.44\pm0.01)\%$ & $36.6\pm0.2$ & $0.39532\pm0.00008$\\
      	& M/40 & $(0.123\pm0.004)\%$ & $(0.832\pm0.003)\%$ & $36.7\pm0.2$ & $0.39765\pm0.00003$\\
	& extrapolation & $(0.118\pm0.002)\%$ & $(0.546\pm0.002)\%$ & $36.7\pm0.2$ & $0.39876\pm0.00002$\\
MS++  	& M/25 & $(0.065\pm0.004)\%$ & $(1.038\pm0.008)\%$ & $39.5\pm0.2$ & $0.19817\pm0.00003$\\
      	&   M/30 & $(0.063\pm0.003)\%$ & $(0.7698\pm0.0040)\%$ & $39.6\pm0.2$ & $0.19868\pm0.00003$\\
      	& M/35 & $(0.062\pm0.002)\%$ & $(0.6567\pm0.0069)\%$ & $39.5\pm0.2$ & $0.19891\pm0.00003$\\
	& extrapolation & $(0.060\pm0.002)\%$ & $(0.4986\pm0.0086)\%$ & $39.5\pm0.2$& $0.19920\pm0.00007$\\
0S++  	& M/22.5 & $0.0616\pm0.0032$ & $0$ & $40.4\pm0.2$ & $0$\\
      	& M/25 & $(0.0595\pm0.0025)\%$ &  $0$ & $40.4\pm0.2$ & $0$\\
      	& M/30 & $(0.057\pm0.002)\%$ & $0$ & $40.4\pm0.2$ & $0$\\
	& extrapolation & $(0.054\pm0.002)\%$ & $0$ & $40.4\pm0.2$ & $0$
\end{tabular}
\end{ruledtabular}
\end{table}
\end{widetext}

\begin{figure}
\begin{center}
\includegraphics[width=3.3in]{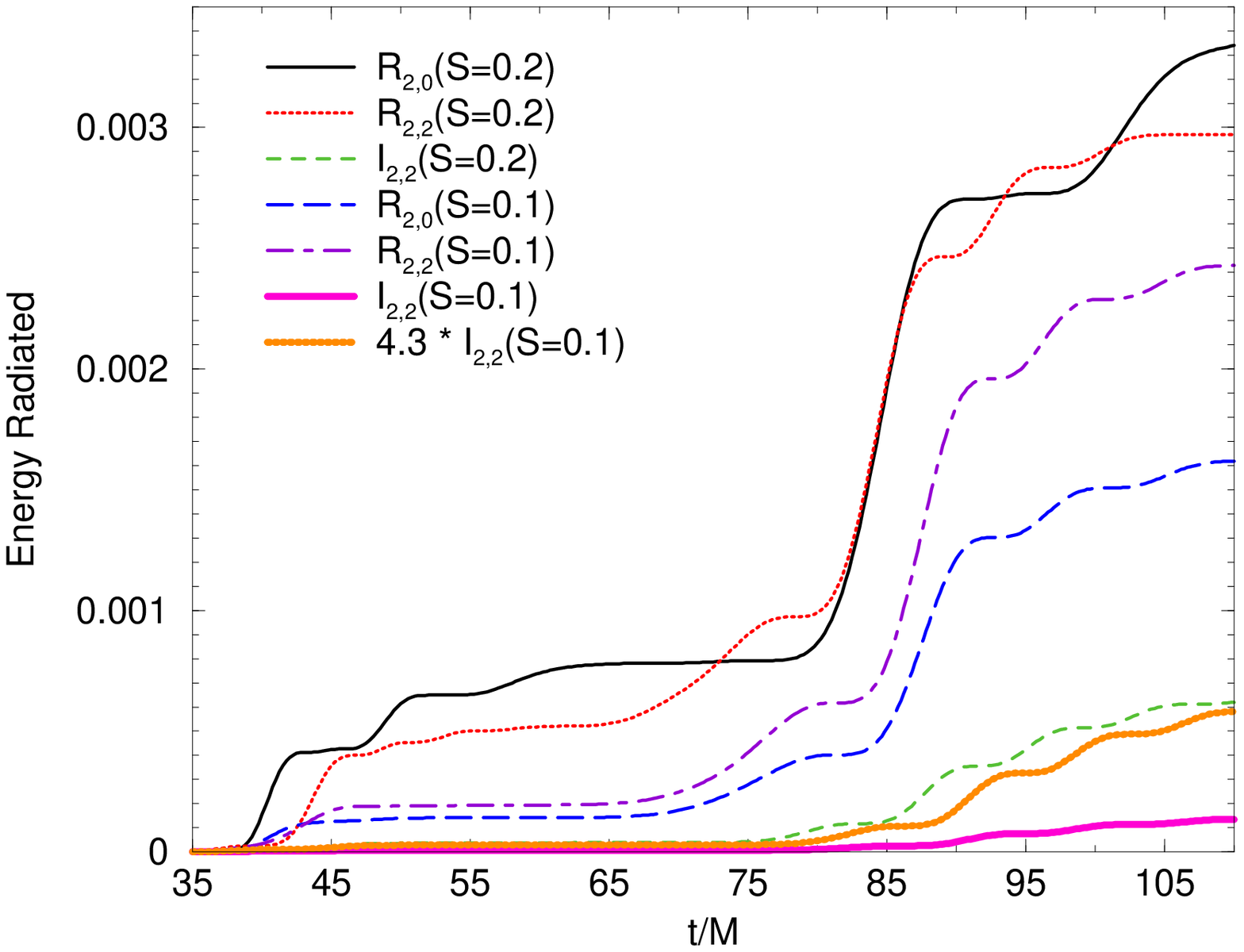}
\caption{The gravitational energy radiated up to time $t$ as measured
by an observer at $r=35M$ for the $\ell=2$ modes of the HS++ and MS++ configurations
(for our configurations $R_{2\, 2} = R_{2\, -2}$, and $I_{2\, 2} = I_{2\, -2}$). The 
energy from the spurious initial burst leads to the first plateaus 
centered at $t=60M$.
These initial bursts of energy, while significant, are dominated by
the total emitted
radiation. We multiply $I_{2\,\pm2}$ from the MS++ configuration ($S=0.1M^2$)
by $4.3$ to demonstrate that quadratic (plus smaller higher-order terms)
dependence of  $I_{2\,\pm2}$ on $S$.
}
\label{fig:vhs_ls_l2_compare}
\end{center}
\end{figure}

An analysis of the radiated energy versus spin is complicated by the
fact that the binary radiates energy in the zero-spin configuration
as well. After subtracting the zero-spin radiated energy we find that
the HS++ configuration radiates $0.064\pm0.003 M$ more energy than the zero-spin
configuration, while the MS++ configuration only radiates
$0.006\pm0.002 M$ more. Thus the HS++ configuration radiates
$12.2\pm4.6$ times as much as the MS++ configuration (after subtracting off the
zero-spin radiated energy). From these
results it appears that the radiated energy scales with $j_{0}^3$
or $j_{0}^4$. A close-limit analysis (See Eq.~(\ref{Ecl})) also indicates that there is a
small $j_{0}^2$ contribution as well. Although not obvious in the
total radiated energy, this quadratic dependence on $J$ can be seen in
the imaginary part of the $(\ell=2,m=\pm2)$ components of the radiated
energy $E(t)$ (i.e.\ the amount of radiation emitted up to time $t$),
where $E(t) = \sum_{\ell, m} (R_{\ell\,m}(t) + I_{\ell\,m}(t))$,
\begin{eqnarray}
  R_{\ell\,m}(t) = \frac{1}{4 \pi} \int_0^t Re[N_{\ell\,m}(\tau)]^2 d\tau, \\
  I_{\ell\,m}(t) = \frac{1}{4 \pi} \int_0^t Im[N_{\ell\,m}(\tau)]^2d\tau,
\end{eqnarray}
$N_{\ell\,m}(t) = (r - 2 M)/2  \int_0^t \bar \psi_{4,\,\ell\,m}(\tau)\, d\tau$,
and $\psi_4$ is calculated using the quasi-Kinnersley tetrad method~\cite{Campanelli:2005ia}.
Figure~\ref{fig:vhs_ls_l2_compare} shows $R_{\ell\,m}(t)$ and $I_{\ell\,m}(t)$
 for the $\ell=2$ components
of the waveform for the HS++ and MS++ configurations. Note that
$I_{2\,\pm2}$ (which vanishes in the zero-spin case) is 4.3 times larger
in the HS++ configuration than in the MS++ configuration. This
is consistent with a quadratic
dependence on the spin plus smaller higher-order dependencies. 

Finally, if we fit the radiated energy to the polynomial dependence
$E_0 + E_2 j_0^2 + E_4 j_0^4$ we obtain
\begin{eqnarray}
\label{eq:EradVJ}
\frac{E_{rad}}{M}&=&(0.052\pm0.02)+(0.065\pm0.033)j_0^2 \nonumber\\
&&+(2.08\pm0.21)j_0^4 + O\left(j_0^6\right).
\end{eqnarray}
Note that the coefficient of the quadratic term is two orders of magnitude smaller than the
coefficient of the quartic term (and thus was not apparent in our initial
analysis above).  There are only even terms in Eq.~(\ref{eq:EradVJ}) because the $E_{rad}$ cannot
depend on the sign of $j_0$ (i.e.\ $E_{rad}$ must be the same for configurations with the spins
aligned on the $+z$-axis and with the spins aligned on the $-z$-axis).

The Bowen-York initial data family contains `spurious' radiation
that does not represent any physical history of the
binary.  In addition, the spinning case does not have the Kerr limit when
the two black holes are far apart. (For a data set with the right Kerr
limit see \cite{Dain:2000hk,Dain:2001iw}.) While this extra radiation
is of relatively small magnitude in orbiting black-hole binaries (See
~\cite{Campanelli:2006uy}), it is relevant in the highly spinning
head-on collisions considered here.  This can be observed in
Fig.~\ref{fig:vhs_ls_l2_compare}. The first plateau in the radiated
energy ($t=40M - 70M$) for the observer location at $r=35M$
is due to the radiated `spurious' energy. This `spurious' energy contributes up to
$25\%$ of the total energy radiated of the HS++ configuration, but strongly decreases
for the MS++ and 0S++ configurations.

One can use the lowest relevant post-Newtonian expansion to
support (qualitatively) our results. 
A look at the Post-Newtonian equations of motion~\cite{Kidder:1995zr}
shows that the leading spin-orbit interaction is of 1.5PN order, while
the leading spin-spin interaction is of 2PN order. Kidder~\cite{Kidder:1995zr}
gives the radiated energy and momenta of the point-particle
spin-orbit and spin-spin couplings up to 2PN order. The relevant
equations for our nearly head-on collisions are:
\begin{subequations}\label{EdotPN}
\begin{eqnarray}
\dot{E}_N&=&-\frac{8 \bar m^4}{15r^4}v_r^2,\\
\dot{E}_{PN}&=&-\frac{2\bar m^4}{105r^4}\left\{-110v_r^4+452\frac{\bar m}{r}v_r^2\right\},\\
\dot{E}_{SO}&=&0,\quad(\vec{L}_N=0),\\
\dot{E}_{SS}&=&\frac{32 \bar m^2}{5r^6}(\vec{S}_1\cdot\vec{S}_2)v_r^2,
\end{eqnarray}
\end{subequations}
where $r$ is the binary separation, $2 \bar m=m_1+m_2$, $\vec v_{r}$ is the relative velocity, $N$ denotes Newtonian terms,
PN denotes (non-spin-orbit and non-spin-spin) post-Newtonian terms,
SO denotes spin-orbit coupling terms, and SS denotes
spin-spin coupling terms. Note that the lowest-order spin-correction to the radiated energy
is quadratic in $S$. 
Likewise we can obtain the radiated angular momentum
\begin{subequations}\label{JdotPN}
\begin{eqnarray}
\dot{\vec{J}}_N&=&0,\quad(\vec{L}_N=0),\\
\dot{\vec{J}}_{PN}&=&0,\quad(\vec{L}_N=0),\\
\dot{\vec{J}}_{SS}&=&0\quad(\vec{a}_N\sim\vec{a}_{SS}\sim\hat{r}),\\
\dot{\vec{J}}_{SO}^{++}&=&-\frac{4 \bar m^2}{15r^3}\,\vec{v}_r\times(\vec{v}_r\times
(\vec{S}_1+\vec{S}_2))\nonumber\\
&&\left\{v_r^2+\frac{2 \bar m}{r}+\frac{1}{v_r^2}\frac{4 \bar m^2}{r^2}\right\}\label{JdotPNSO},
\end{eqnarray}
\end{subequations}
which displays the leading linear dependence on the spin,
and the radiated linear momentum
\begin{subequations}\label{PdotPN}
\begin{eqnarray}
\dot{\vec{P}}_N&=&0,\quad(\delta m=m_1-m_2=0)\\
\dot{\vec{P}}_{PN}&=&0,\quad(\delta m=m_1-m_2=0)\\
\dot{\vec{P}}_{SS}&=&0\quad{\rm (at\ this\ order)}\\ 
\dot{\vec{P}}_{SO}^{++}&=&-\frac{8 \bar m^3}{15r^5}\,{v}_r(\vec{v}_r\,\times\vec{\Delta})=0,\nonumber\\
&&(\vec{\Delta}=2(\vec{S}_1-\vec{S}_2)=0).\label{PdotPNSO}
\end{eqnarray}
\end{subequations}

Note that equation (\ref{PdotPNSO}) gives a non-vanishing
linear momentum radiated for anti-parallel spins.
The spin-orbit interaction is the only one that reverses signs
when spins change signs. This explains the different stability and
radiation properties of parallel and anti-parallel configurations.
We plan to report the results of this spin up - spin
down case leading to radiated {\it linear} momentum in a forthcoming paper.

These spinning-black-hole-binary scenarios were also studied in the complementary regime of the close
limit approximation by Gleiser and Dom\'{i}nguez
\cite{Gleiser:2001in}. They examined the behavior of  non-boosted, slowly rotating holes to second
perturbative order. In that approximation the second-order terms
proportional to the spin and the square of the proper distance, $l$,
give rise to radiation of the angular momentum, $J$, proportional to
the spin
\begin{equation}\label{Jcl}
J_{rad}\approx0.0023\, (2S)\,(l/M)^4.
\end{equation}
The analysis also predicts that the radiation of energy can be expanded as
\begin{eqnarray}\label{Ecl}
\frac{E_{rad}}{M}&\approx&7.8\times10^{-4}(2S/M^2)^4+9.8\times10^{-5}\,(l/M)^4\nonumber\\
&-&2.8\times10^{-5}\,(2S/M^2)^2(l/M)^2\nonumber\\
&+&1.3\times10^{-3}(2S/M^2)^2(l/M)^4,
\end{eqnarray}
The first term, being proportional only to $S^4$, is generated by the
radiation content of the BY initial data. The second term, proportional
to $l^4$, is the radiation for two non-spinning masses in a head-on
collision. The third term is second-perturbative-order spin-orbit
coupling term, and is the one that leads to
the radiation of angular momentum.

\section{Discussion of Results}\label{Sec:discussion}

We have studied the spin-up of initially non-spinning and initially corotating
black-hole binaries during the last
stages of the inspiral orbital motion and found that it is much
smaller than that needed to achieve tidal locking.
This means that the time scale of tidal effects to spin up the black holes
is much longer than the time scale for gravitational radiation to
drive the merger (and hence increase the orbital frequency).
This
also implies that in an astrophysical scenario, black holes with low
spin merge with essentially the same spin magnitude they had when the binary
formed (note that Eq.~(\ref{Sdot}) implies that $|\dot{\vec{S}}|=0$).
We also note that it is this short timescale for gravitational radiation
during the last few orbits that is responsible for the absence of an
ISCO (innermost stable circular orbit) when the two holes have comparable
masses. The concept of the ISCO arises naturally in the extreme mass ratio
limit, where there is a cutoff between stable and unstable orbits. In this
limit, the instability of orbits inside the ISCO is not driven by gravitational
radiation, but by non-radiative terms in the particle's equation of motion. 
Thus the particle very slowly inspirals due to the emission of weak gravitational
radiation, and then plunges when it reaches the ISCO.
On the other hand, in the comparable mass regime, there is no such
cutoff because gravitational radiation dominates the dynamics during the
entire late-time inspiral and plunge phases.

\begin{figure}
\begin{center}
\includegraphics[width=3.3in]{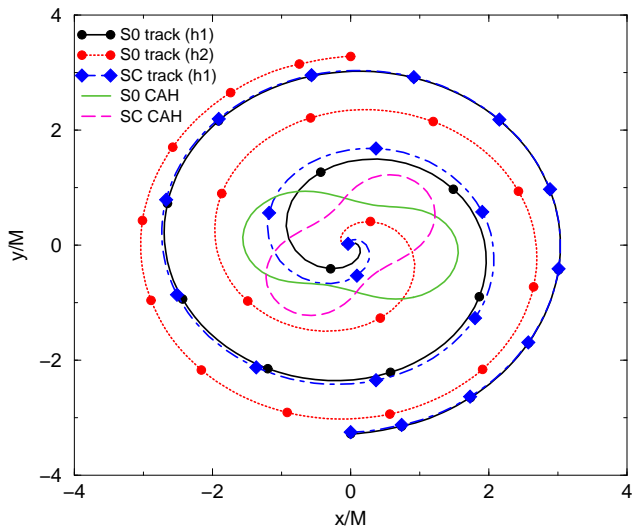}
\caption{The horizon tracks for the S0 and SC configurations
with tick marks every $10M$ of evolution
(we display only one of the SC tracks for clarity). Note that the
S0 and SC tracks agree well for the first $100M$, and that after
$100M$ the SC tracks show a tighter spiral and orbit longer
before merging. The S0 configuration completes slightly less than
1.75 orbits, while the SC configuration completes slightly more.}
\label{fig:tracks}
\end{center}
\end{figure}

In Fig.~\ref{fig:tracks} we show the tracks of the horizon centroids in
the S0 and SC configurations. Although coordinate dependent, the tracks
help illustrate the lack of an ISCO in the comparable mass regime. Rather
than showing an abrupt change near the last orbit, the tracks show a smooth
increase in plunge velocity over the entire evolution. Note that the 
SC track is a tighter spiral than the S0 track, which is consistent with
the results found previously by the authors~\cite{Campanelli:2006uy}.
Additional horizon tracks for multi-orbit non-spinning binaries
can be found in Refs.~\cite{Baker:2006yw, Campanelli:2006gf},
while tracks for highly-spinning
binaries can be found in Ref.~\cite{Campanelli:2006uy}.

After the merger, the final remnant black hole formed from the S0 configuration
has a specific spin of $a/M_{H}=0.688\pm0.001$ (where $M_{H}$ is the
remnant horizon mass) while remnant formed from the
SC configuration has a specific spin of $a/M_{H}=0.717\pm0.001$. Using these
values in conjunction those found in Ref.~\cite{Campanelli:2006uy}
(see Table~\ref{tab:spindat}), we
find (from a least-squares fit) that the specific spin of the final
remnant horizon $(a/M_{H})|_R$ varies 
with the initial specific spins of the two (equal-mass, equal-spin) black 
holes $(a/m)|_{I}$ according to:
\begin{eqnarray}
  (a/M_{H})|_{R} &=& 0.6879 + 0.2952 \left((a/m)|_{I}\right) \nonumber \\
             &-&0.0374 \left((a/m)|_{I}\right)^2,
\end{eqnarray}
with an error of $\pm0.001$ in the range $(-0.757 < (a/m)|_{I} < 0.757)$.
From this fit we extrapolate that the remnant horizon will have a specific spin
no smaller than $0.355$ (for the case of initially-maximal, anti-aligned
spins) and no larger than
$a/M_{H} = 0.946$ (for the case of initially-maximal, aligned spins),
which supports our assertion in~\cite{Campanelli:2006uy} that
the cosmic censorship conjecture is not violated by the merger of
two black holes.
A least-squares fit of the energy radiated versus the initial
individual spins yields
\begin{eqnarray}
  \frac{E_{rad}}{M} &=& 0.0348 + 0.0297 \left((a/m)|_{I}\right) +\nonumber \\
   && 0.0170 \left((a/m)|_{I}\right)^2,
\end{eqnarray}
with an error of $\pm0.001$ in the range $(-0.757 < (a/m)|_{I} < 0.757)$.
Based on this fit we extrapolate that a quasi-circular binary
will radiated no more than $8.1\%$ of its mass during the final
few orbits and merger.

\begin{table}
\caption{The measured remnant horizon specific spin 
$(a/M_{H})|_R$ and energy radiated $E_{rad}$, as well as 
the predicted remnant horizon specific spin
$(a/M_{H})|_{pred}$ and energy radiated $E_{rad}/M|_{pred}$ (based on a least-squares fit) for quasi-circular,
equal-mass, equal-spin binaries with initial specific spins
$(a/m)|_I$.}
\begin{ruledtabular}
\begin{tabular}{lllll}\label{tab:spindat}
$(a/m)|_I$ & $(a/M_H)|_R$ & $(a/M_{H})|_{pred}$ & $E_{rad}/M$ & $E_{rad}/M|_{pred}$\\\hline
$-0.757$ & $0.443\pm0.001$ & $0.4430$ & $(2.2\pm.01)\%$& $2.20\%$\\
$0.000$& $0.688\pm0.001$ & $0.6878$&$(3.5\pm0.1)\%$ &$3.48\%$\\
$0.1001$& $0.717\pm0.001$ & $0.7169$&$(3.8\pm0.1)\%$&$3.79\%$\\
$0.757$ & $0.890\pm0.001$ & $0.8900 $&$(6.7\pm0.2)\%$ &$6.70\%$\\
\hline
$-1.0$ & $***$ & $0.355$& $***$ &$2.2\%$\\
$+1.0$ & $***$ & $0.946$& $***$ &$8.1\%$\\
\end{tabular}
\end{ruledtabular}
\end{table}

We have also found that the tidal coupling in black-hole binaries responsible
for the transfers of angular momentum from the individual spins to
orbital is relatively small. Nevertheless,
we have been able to compute this transfer with sufficient accuracy
to model its dependence on the initial spin. 
We find that the radiated angular momentum, which is a measure of the
angular momentum transferred from spin to orbit, has a linear-plus-cubic
dependence on the initial spins; in qualitative agreement both with
the post-Newtonian (see Eq.~(\ref{JdotPNSO})) and close-limit (see Eq.~(\ref{Jcl})) predictions.
The radiated energy has a quadratic-plus-quartic dependence on the
initial spin; again, in qualitative agreement with the post-Newtonian
(see Eqs.~(\ref{EdotPN})) and close-limit (see Eq.~(\ref{Ecl})) predictions.

Unfortunately the speculations of Price and Whelan~\cite{Price:2001un} suggesting this spin-up
effect might be strong, and even dominant, are not confirmed by our computations.
We see neither the possibility of a binary stalling by this effect nor 
the dominance of the $\ell=3$ mode suggested in 
Ref.~\cite{Price:2001un}.

\appendix
\section{Isolated horizon spin and horizon circumferences}
\label{app:isolated_v_cir}
The most common technique for evaluating the spin of a numerically
evolved black hole is to use the ratio of the polar to equatorial
circumferences~\cite{Alcubierre:2004hr}. For a Kerr hole this ratio
$C_{r} = C_{p} / C_{eq}$ is given by
\begin{equation}
C_r = \frac{1+\sqrt{1-(a/m)^2}}{\pi} \,
        E \left( -\frac{(a/m)^2}{(1+\sqrt{1-(a/m)^2})^2}
\right),
\label{eq:Cr-exact}
\end{equation}
where $a/m$ ($m$ being the horizon mass) is the specific spin and $E(x)$ is the complete elliptic
integral of the second kind. After a common horizon forms, $C_r$ displays
quasinormal ringing with a constant offset. To obtain the specific spin in
finite time one uses a non-linear least-squares fit
of $C_r$ versus time to $C_r = C_0 + C_1 \exp[-C_2 t/M] \sin[C_3 t/M+C_4]$
to obtain $C_0,\cdots,C_4$ ($C_0$ being the asymptotic value of
$C_r$), and then $a/m$ is obtained by inverting Eq.~(\ref{eq:Cr-exact}).
The authors have used this technique to obtain
highly-accurate evaluations of the remnant spin in
Refs.~\cite{Campanelli:2005dd,Campanelli:2006gf,Campanelli:2006uy}.
However, in the pre-merger case this technique
is not accurate. There are two reasons for this. First, one cannot
separate the large oscillations in $C_r$ from the secular growth
because the behavior of the oscillations is not know a-priori. 
Thus, there is a large error in the calculated spin. Second, the
distortions on the horizon that lead to $C_{r} \neq 1$ are due to both
the spin as well as tidal distortion of the horizon, and when the black
holes are close, the tidal distortion dominate. Thus even if one can
obtain an accurate measurement of the secular part of $C_r$, one would
then need to disentangle the spin component of $C_r$ from the tidal
distortion component. Finally, the horizon circumference calculations
are not coordinate invariant and lead to errors when the coordinate
are distorted (i.e.\ when the coordinate circles used in the
calculations do not correspond to the correct coordinate circles in
Boyer-Lindquist coordinates).

\begin{figure}
\begin{center}
\includegraphics[width=3.3in]{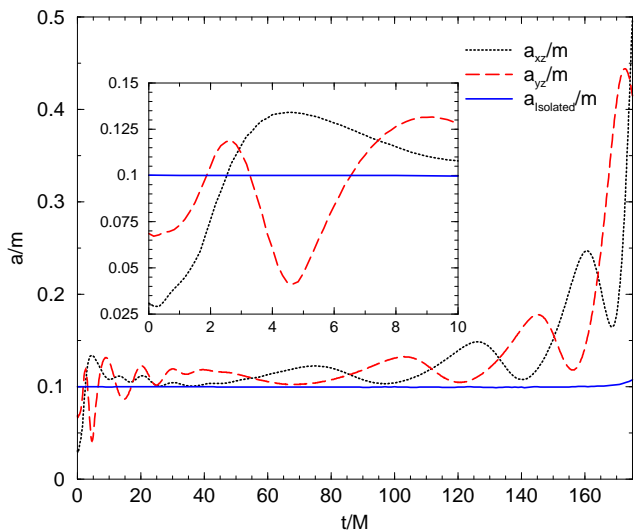}
\caption{The specific spin of the individual horizons for the
SC orbiting binary configuration as calculated using the
horizon circumferences and the isolated horizon technique.
Here $a_{xz}$ is obtained from $C_{xz}/C_{eq}$ via Eq.~(\ref{eq:Cr-exact}), $a_{yz}$ from
$C_{yz}/C_{eq}$. Note the large oscillations and secular growth in
$a_{xz}$ and $a_{yz}$ compared to the much more accurate isolated horizon spin.
}
\label{fig:kill_cir_compare}
\end{center}
\end{figure}

In Fig.~\ref{fig:kill_cir_compare} we show $a/m$ for the SC
orbiting-binary configuration calculated using the isolated horizon
method and the circumference method. In the plot we show $a/m$
calculated using the ratios of the $xz$ and $yz$ polar
circumferences to the equatorial (i.e.\ $xy$) circumference, as well
$a/m$ derived from the isolated horizon technique.
Note that
the initial values for the spin 0.03 and 0.07 for the two
circumference-based calculations are significantly smaller
than the expected $0.1$, while
the isolated horizon technique gives the correct value. The difference
between the spin calculated from the two circumferences is a measure
of the error. In this case that error is larger than the effect we
want to measure. Note also that the spin derived from the
circumferences shows a much stronger increase with time. This strong
increase is driven by the increasing tidal distortion of the horizon
(as can be inferred from the increase in amplitude of the oscillations).

\acknowledgments
We thank Bernard Kelly for careful reading of this text.
We thank
Erik Schnetter for providing the Cactus thorns to implement
Pi-symmetry boundary conditions,  Marcus Ansorg for providing the
{\sc TwoPunctures} thorn, and Badri Krishnan for suggesting the comparison
 between the
isolated horizon technique and the horizon circumference technique.
We gratefully acknowledge the support
of the NASA Center for Gravitational Wave Astronomy at University of
Texas at Brownsville (NAG5-13396) and the NSF for financial support
from grants PHY-0140326 and PHY-0354867. Simulations were
performed on the 70-node `Funes' cluster at UTB, on the `Lonestar2'
supercomputer at TACC, and on the `Tungsten' supercomputer at NCSA.

\bibliographystyle{apsrev}
\bibliography{../../Lazarus/bibtex/references}

\end{document}